\documentclass{mn2e}

\input psfig

\title[Variability of black hole accretion]
        {The variability of accretion onto Schwarzschild black holes from 
        turbulent magnetized discs}
        
\author[P.J. Armitage \& C.S. Reynolds]
       {Philip J. Armitage$^{1,2}$\thanks{email: {\tt pja@jilau1.colorado.edu}}  
       and Christopher S. Reynolds$^3$\thanks{email: {\tt chris@astro.umd.edu}} \\
        $^1$JILA, University of Colorado, 440 UCB, Boulder CO 80309-0440, USA \\
        $^2$Department of Astrophysical and Planetary Sciences, University of Colorado, 
            Boulder CO 80309-0391, USA \\ 
        $^3$Department of Astronomy, University of Maryland, College Park MD 20742, USA \\}     
 
\begin{document}

\maketitle

\begin{abstract}
We use global magnetohydrodynamic simulations, in a pseudo-Newtonian
potential, to investigate the temporal variability of accretion discs
around Schwarzschild black holes. We use the vertically-averaged
magnetic stress in the simulated disc as a proxy for the rest-frame
dissipation, and compute the observed emission by folding this through
the transfer function describing the relativistic beaming, light
bending and time delays near a non-rotating black hole. The temporal 
power spectrum of the predicted emission from individual annuli in 
the disc is described by a broken power law, with indices of $\approx -3.5$ 
at high frequency and $\approx 0$ to $-1$ at low frequency. Integrated 
over the disc, the power spectrum is approximated by a single power law 
with an index of $-2$. Increasing inclination boosts the relative power at 
frequencies around $\approx 0.3 f_{\rm ms}$, where $f_{\rm ms}$ is 
the orbital frequency at the marginally stable orbit, but no 
evidence is found for sharp quasi-periodic oscillations in 
the lightcurve. Assuming that fluorescent iron line emission
locally tracks the continuum flux, we compute simulated broad iron
line profiles. We find that relativistic beaming of the
non-axisymmetric emission profile, induced by turbulence, produces
high-amplitude variability in the iron line profile.  We show that
this substructure within the broad iron line profile can survive
averaging over a number of orbital periods, and discuss the origin of 
the anomalous X-ray spectral features, recently reported by Turner et
al. (2002) for the Seyfert galaxy NGC~3516, in the context of turbulent 
disc models.
\end{abstract}


\section{Introduction}
A major aim of the astrophysical study of black holes is to determine
the fundamental parameters of a rotating black hole -- the mass $M$
and spin parameter $a$ -- from astronomical observations. For Galactic
black hole candidates such as Cygnus X-1, GRO J1655-40 and GRS
1915+105, the temporal power spectrum holds promise as a route to
achieving this goal (van der Klis 2000). The variability of these
sources is conventionally described as the superposition of a
broadband noise component with one or more quasi-periodic oscillations
(QPOs), whose frequencies presumably encode information about the
inner accretion flow and black hole properties. For AGN, where the
long dynamical times associated with supermassive black holes hamper
similar studies, the main focus is on inferring black hole properties
from the profiles of relativistically broadened iron emission lines
arising from near the marginally stable orbit (Fabian et al. 1989;
Tanaka et al. 1995; Nandra et al. 1997; Wilms et al. 2001; Lee et
al. 2002). Iron line profiles themselves are highly variable (Iwasawa
et al. 1996).

For accreting black holes, and for other relativistically compact
objects, the observed lightcurve arises from the modulation of
rest-frame variability (i.e. variability that would be detected by an
observer orbiting with the disc gas) by photon propagation effects.
Since there is now a consensus that magnetorotational instabilities
(MRI, Balbus \& Hawley 1991; Balbus \& Hawley 1998) provide a source
of angular momentum transport in well ionized discs, an unavoidable
source of rest-frame variability is the fluctuating dissipation
inherent to magnetohydrodynamic (MHD) disc turbulence. Simulations of
the MRI show that the high frequency temporal power spectra of the
mass accretion rate and magnetic stress display steep power-law
spectra (Kawaguchi et al. 2000; Hawley \& Krolik 2001, 2002), and
suggest that the broadband noise component of observed power density
spectra may be identified with the temporal fluctuations of MHD disc
turbulence. Spatial fluctuations in the dissipation may also play an
important role. For a disc inclined to the line of sight, beaming
enhances the emission from the approaching side of the disc relative
to that from the receeding side. The combination of this photon
propagation effect with any source of persistent non-axisymmetric
dissipation in the inner regions of the disc could lead to high
frequency QPOs in the lightcurve (e.g. Abramowicz et al. 1991; Karas 1999).

In this paper, we use non-relativistic global MHD simulations of the
MRI (Armitage 1998; Hawley 2000; Arlt \& Rudiger 2001) to model the
rest-frame variability of accretion onto a Schwarzschild black hole.
A pseudo-Newtonian potential (Paczynski \& Wiita 1980) is used to
describe the gravitational potential in the vicinity of such a black
hole.  We extend recent three-dimensional numerical studies of black
hole accretion (Hawley 2000; Hawley \& Krolik 2001; Armitage, Reynolds
\& Chiang 2001; Hawley 2001; Reynolds \& Armitage 2001; Hawley, 
Balbus \& Stone 2001; Machida, Matsumoto \&
Mineshige 2001; ; Hawley \& Krolik 2002; Igumenshchev \& Narayan 
2002; Machida \& Matsumoto 2003; Igumenshchev, Narayan \& Abramowicz 
2003) by taking full account of the distinction between the rest-frame
disc emission and that seen by a distant observer. We use these
simulations to study both the predicted power spectra from black holes
accreting via a geometrically thin disc, and the expected variability
of the iron line emission that has been observed in the X-ray spectra
of Seyfert galaxies (Tanaka et al. 1995; Wilms et al 2001; Lee et al.
2002) and Galactic Black Hole Candidates (Balucinska-Church \& Church
2000; Miller et al. 2002).  The details of the MHD simulations are
presented in Section~2, while Section~3 describes our computation of
the observed continuum and line emission taking into account
relativistic photon propagation.  Predicted temporal power spectra are presented in
Section~4.  In Section~5, we explore our predictions for variability
of the broad iron line.  It is suggested that the non-axisymmetric
structure discussed in this paper may already have been observed in
{\it Chandra} High Energy Transmission Grating (HETG) data for the
Seyfert galaxy NGC~3516, and we proceed to quantify the nature 
of the variability that might be observed in the future, using higher 
signal-to-noise {\it Constellation-X} data. Our results, predictions
and speculations are summarized in Section~6.

\section{The MHD simulations}
We model the accretion disc around a non-rotating black hole using
ideal, Newtonian MHD. The fluid is assumed to be isothermal, with
sound speed $c_s$, so the equations to be solved are,
\begin{eqnarray}
 { {\partial \rho} \over {\partial t} } + \nabla \cdot ( \rho {\bf v} ) & = & 0 \\
 { {\partial {\bf S}} \over {\partial t} } + \nabla \cdot ( {\bf S} {\bf v} ) & = & 
 - \nabla P - \rho \nabla \Phi + {\bf J} \times {\bf B} \\
 { {\partial {\bf B}} \over {\partial t} } & = & \nabla \times ({\bf v} \times {\bf B}) \\
 P & = & \rho c_s^2,
\end{eqnarray} 
where ${\bf S} = \rho {\bf v}$ is the momentum field and the other
symbols have their usual meanings. These equations are solved using
the ZEUS MHD code (Stone \& Norman 1992a, 1992b), operating in
cylindrical polar co-ordinates $(z,r,\phi)$. ZEUS is a fixed-grid,
time-explicit Eulerian code which uses an artificial viscosity to
handle shocks, and a combination of constrained transport (Evans \&
Hawley 1988) and the method of characteristics (Stone \& Norman 1992b)
to evolve the magnetic fields.  When operated using van Leer
advection, as in this work, it is formally of second order accuracy.

A pseudo-Newtonian gravitational potential (Paczynski \& Wiita 1980)
is used to mimic the effects of general relativity near a non-rotating
black hole of mass $M$,
\begin{equation}
 \Phi = - { {GM} \over {r - r_g} },
\end{equation} 
where $r_g = 2 GM / c^2$. This potential reproduces the presence of a
marginally stable circular orbit at $r_{\rm ms} = 6GM / c^2$. For
computational reasons, we also omit the vertical component of the
gravitational force.  This `cylindrical disc' approximation has been
used in several previous studies (Hawley, Gammie \& Balbus 1995; Armitage 1998; Reynolds \& Armitage
2001; Hawley 2001; Steinacker \& Papaloizou 2002), and greatly reduces
the computational cost of global disc simulations. Of course, it also
precludes study of physical effects such as the Parker instability and
the development of a disc corona (e.g. Miller \& Stone 2000).

To investigate variability caused by the combination of a turbulent
disc with relativistic beaming effects, it is necessary to model all
$2 \pi$ of the disc in azimuth. Apart from this (trivial) extension,
our numerical setup is the same as that which we have used previously 
(Armitage, Reynolds \& Chiang 2001; Reynolds \& Armitage 2001). A
hydrodynamically stable disc flow, threaded by a weak vertical
magnetic field, is set up as the initial state of the simulations. We
begin by defining a Gaussian surface density profile, centred at
4~$r_{\rm ms}$, where $r_{\rm ms}$ is the radius of the marginally
stable circular orbit, with a width of 2.66~$r_{\rm ms}$. This disc,
initialized with Keplerian angular velocity and zero magnetic fields,
is run with the code in 1D mode until a numerical equilibrium is
attained. We then add a weak vertical magnetic field (ratio of gas to
magnetic pressure $\beta = 10^4$) at all radii where the density
exceeds a threshold value (set to be approximately 10 percent of the
maximum density), and continue the run in 3D. The boundary conditions
are periodic in both $z$ and $\phi$, and set to outflow at the radial
boundaries. `Outflow' boundary conditions are implemented in ZEUS by
setting the values of all physical quantities in boundary zones to be
equal to their values in the active zones at the edge of the grid.

Two simulations were run, differing only in the adopted sound
speed. One run had a ratio of the sound speed to the Keplerian
velocity at $r_{\rm ms}$ of $c_s / v_\phi = 0.061$, while a 
second run had $c_s / v_\phi = 0.041$ at $r_{\rm ms}$. We will 
refer to these as the `hot' and `cold' run respectively, though 
the actual temperature differences are modest. Both simulations 
were evolved for a total of 140 orbital periods at $r_{\rm ms}$, 
in identical computational domains,
\begin{eqnarray} 
 -0.7 r_{\rm g} < & z & < 0.7 r_{\rm g} \\
 0.7 r_{\rm ms} < & r & < 8 r_{\rm ms} \\
 0 < & \phi & < 2 \pi,
\end{eqnarray} 
with 48 grid points in $z$, 192 grid points in $r$, and 288 grid
points in $\phi$. A non-uniform radial grid, with mesh points spaced
such that $r_{i+1} = (1 + \delta) r_i$, with $\delta$ a constant, is
employed, so that $\Delta r / \Delta \phi$ is constant at all
radii. The inner radial boundary was placed as close to the marginally
stable orbit as possible (to maximise the timestep which is almost
always determined by the azimuthal velocity at $r_{\rm in}$), but far
enough within $r_{\rm ms}$ that the flow across the boundary was both
supersonic and super-Alfvenic. As has been noted by Igumenshchev et al. (2003), 
this inner boundary condition becomes inappropriate if large net amounts of 
magnetic flux are able to be dragged inward by the accretion flow. Here, 
we make the assumption -- justified in part by the results of Lubow, Papaloizou \& 
Pringle (1994) -- that extensive field dragging does not occur for thin discs.

\subsection{Properties of the simulations}
The basic properties of the MHD disc turbulence seen in the simulations
are, unsurprisingly, very similar to those reported previously
(i.e. although we need a full $2\pi$ in azimuth to study variability,
smaller azimuthal domains suffice for most other purposes). Here, we
summarize a few features that are pertinent to the subsequent analysis.

\begin{figure}
 \psfig{figure=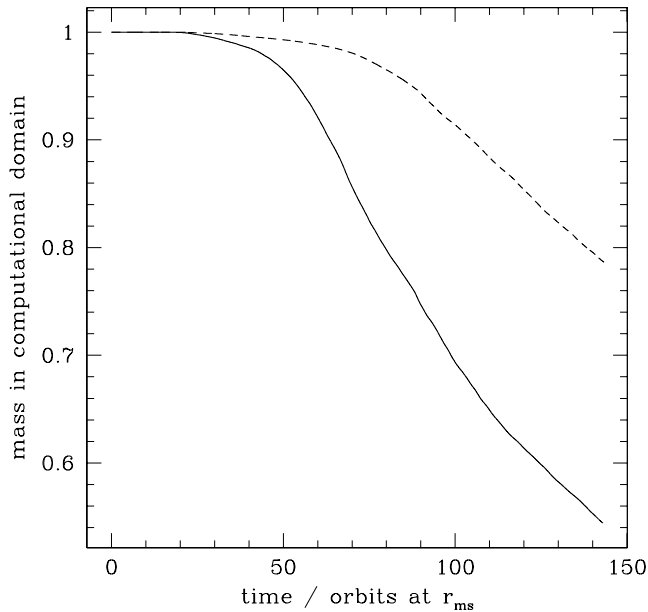,width=3.5truein,height=3.5truein}
 \caption{The disc mass as a function of time for the `hot' (solid 
        curve and `cold' (dashed curve) runs, shown as a fraction 
        of the initial disc mass. Only the fully turbulent portion 
        of the run with $t > 50$, where $t$ is measured in units of 
        the orbital period at $r_{\rm ms}$, is used for analysis of 
        the variability.}
 \label{fig_mass}
\end{figure}

Figure~\ref{fig_mass} shows the total mass in the computational domain
as a function of time for the hot and cool runs
respectively. Significant accretion commences after around 30 orbits
of evolution at the marginally stable orbit. For the hot run, the
accretion rate is very roughly constant after around 50 orbits of
evolution, and it is this portion of the simulation which we analyse
for variability. Note, however, that there remains a secular trend in
the accretion rate even over this latter portion (initially
increasing, then decreasing once a significant fraction of the initial
mass has been accreted). Since this is an artefact of the
computational setup, we can only look at variability on substantially
shorter timescales than the run length\footnote{Because we have only 
one realization of each simulation, there are also large {\em uncertainties} 
in our estimation of the temporal power spectrum at the lowest frequencies.
These purely statistical uncertainties would preclude us from making 
statements about variability on timescales comparable to the run length, 
even if the mean accretion rate showed no secular trend.}

\begin{figure}
 \psfig{figure=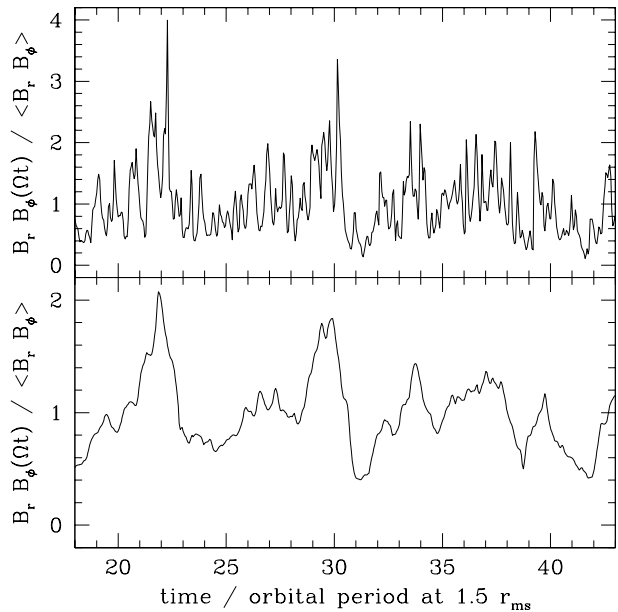,width=3.5truein,height=3.5truein}
 \caption{The vertically integrated magnetic stress for the hot
 disc at a radius of $1.5 \ r_{\rm ms}$, evaluated at a
 single azimuthal point corotating with the local Keplerian angular
 velocity. The upper panel shows the raw measurement, the lower panel
 the same curve smoothed over an interval equal to the local orbital
 period.  The stress is shown normalized to the instantaneous average
 value at the same radius.}  \label{fig_fluc}
\end{figure}

A single persistent hotspot, orbiting near the innermost stable orbit
where relativistic effects are most pronounced, would lead to a strong
periodic feature in the lightcurve (Karas 1999). Such QPOs will be
diluted in more realistic models, first by the existence of patchy
dissipation at {\em all} radii, and second by the limited lifetime of
regions of enhanced dissipation. Figure~\ref{fig_fluc} illustrates the
latter point. We plot the magnetic stress, vertically integrated
through the disc, at a single point corotating with the local
Keplerian orbital velocity.  Results for the hot simulation are
shown, although very similar results are obtained for the cold
case.  Large fluctuations are observed, with the peak stress exceeding
the local mean by a factor that can be as large as $\sim 4$. However,
these are also transient features, with a temporal coherence time of
only a few orbital periods. If, as we assume later, the magnetic
stress provides a proxy for the disc dissipation, it is clear that
`hotspots' arising from MHD disc turbulence will lead to at most broad
QPOs in power spectra of the disc emission.

\begin{figure}
 \psfig{figure=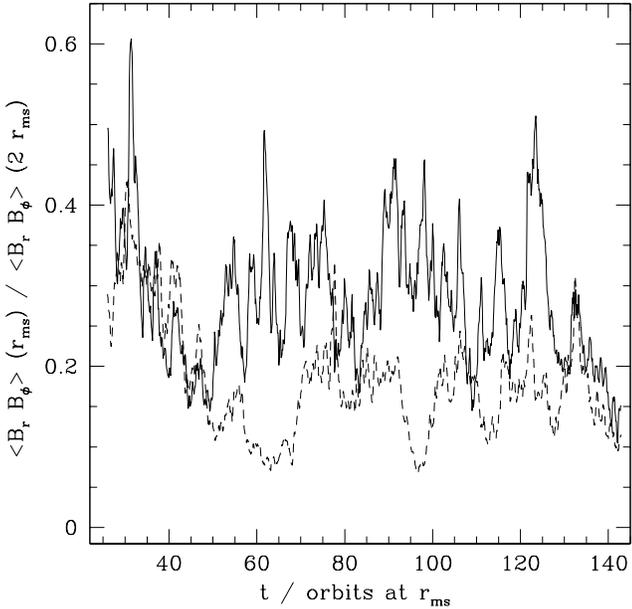,width=3.5truein,height=3.5truein}
 \caption{The magnetic torque at the marginally stable orbit as a 
          fraction of the torque in the disc at a radius 
	  of $2 r_{\rm ms}$ for the simulations 
          with high (solid line) and low (dashed line) sound speed.
          Non-zero stress at $r_{\rm ms}$ becomes relatively more 
          important for the disc structure as the disc temperature 
          increases.}
 \label{fig_torque}
\end{figure}  

Recent analytic (Krolik 1999; Gammie 1999; Agol \& Krolik 2000; 
Afshordi \& Paczynski 2003) and
numerical work (Hawley 2000; Armitage, Reynolds \& Chiang 2001;
Reynolds \& Armitage 2001; Hawley \& Krolik 2001, 2002) has
investigated the possibility that the structure of the inner accretion
disc might be modified by the existence of magnetic coupling to gas in
the plunging region at $r < r_{\rm ms}$. The presence of stress at the
last stable orbit is potentially important for studies of variability,
as a disc with non-zero stress at $r_{\rm ms}$ has a dissipation
profile that is more centrally concentrated than a standard disc (Agol
\& Krolik 2000).  Figure~\ref{fig_torque} shows how important this
effect is for our simulations. We plot the mean stress at $r_{\rm ms}$
as a fraction of the stress at $2 r_{\rm ms}$ (this ratio is adopted
because the changing accretion rate means that the stress at $r_{\rm
ms}$ does not tend to a constant value). At late times, we find that
this ratio fluctuates around a value of around 0.3 for the hot disc
run. There are, however, large fluctuations, which persist over 
timescales that can be as long as 10 orbits at $r_{\rm ms}$. As 
discussed by Papaloizou \& Nelson (2003), and by Winters, Balbus \& 
Hawley (2003), these persistent fluctuations 
are a generic feature of disc MHD turbulence. They are present 
in local (e.g. Brandenburg et al. 1995; Winters, Balbus \& Hawley 2003) 
as well as global simulations (Papaloizou \& Nelson 2003), and 
their existence means that averaging over long timescales is 
needed to recover the mean properties of the flow.

As in previous simulations (Reynolds \& Armitage 2001), there 
is evidence that the contribution of stress at $r_{\rm ms}$ becomes 
relatively less important as the sound speed (or, equivalently, the 
implied disc thickness $h/r$) is reduced.  In our subsequent analysis, 
we compute both raw lightcurves (which include disc dissipation arising 
from stress at $r_{\rm ms}$), and normalized lightcurves which explicitly
fix the form of the radial dissipation profile to match the
predictions of zero-torque boundary condition disc models. Comparing 
these lightcurves allows us to investigate whether the presence of stress 
at the marginally stable orbit has a significant (and potentially 
observable) effect on the predicted 
temporal power spectrum.

\section{Relativistic transfer function}

With a number of approximations, described in Section~4.1, the 
MHD simulations can be used to
define a plausible time-dependent pattern of emission across the disc
plane, $F(r,\phi,t)$. When the region of interest is close to the 
black hole, as it is in this study, one must also account for
general relativistic effects (i.e., beaming, light bending and
time-delays) in order to determine the appearance of the simulated
disc to an observer at infinity.  In this section, we briefly discuss
how we incorporate these effects.

Suppose the observer is viewing the disc from a large distance at an
inclination of $i$ from the normal to the disc plane.  We define
cartesian coordinates $(x,y)$ on the observer's image plane such that
the origin $x=y=0$ lies at the center of the event horizon's image and
the $y$-axis is parallel with the apparent (i.e. projected) normal to
the disc.  Let us consider a photon that is emitted from the disc at
position $(r,\phi)$ with frequency $\nu_{\rm em}$ and is subsequently 
observed, a Schwarzschild time $t$ later, at position $(x,y)$ on the
image plane with frequency $\nu_{\rm obs}$.  Our task is to compute the
functions $x(r,\phi;i)$, $y(r,\phi;i)$, $t(x,y;i)$, and $g(x,y;i)$,
where we have defined the frequency shift factor as
\begin{equation}
g(x,y)=\frac{\nu_{\rm obs}}{\nu_{\rm em}}.
\end{equation}

For a number of inclinations
($i=1^\circ,10^\circ,20^\circ,...,80^\circ$), we integrated null
geodesics (photon paths) through the Schwarzschild metric
\begin{eqnarray}
ds^2&=&-\left({1-\frac{2GM}{c^2r}}\right)dt^2+\left(1-\frac{2GM}{c^2r}\right)^{-1}dr^2 \nonumber \\
 &+& r^2(d\theta^2+\sin^2\theta d\phi^2),
\end{eqnarray}
which exactly describes the space-time geometry around a non-rotating
and uncharged black hole.  Starting from a particular point $(x,y)$ on
the image plane, we integrate photon paths down to the disc plane
$\theta=\pi/2$ using the first integrals of the geodesic equation
(conservation of angular momentum and energy, arising from the $\phi$
and $t$ independence of the metric, respectively), the Carter
constant, and the light-like nature of the path.  For these details,
we refer the reader to Appendix~A of Reynolds et al. (1999).  Turning
points in the photon orbits are identified and treated using the
method of Rauch \& Blandford (1994).  These integrations allow us to
determine the maps $x(r,\phi;i)$, $y(r,\phi;i)$ and $t(x,y;i)$.

The frequency shift factor for a given photon path, $g$, is given by
\begin{equation}
g=\frac{(p_\mu u^\mu)_{\rm disc}}{(p_\mu u^\mu)_{\rm obs}},
\end{equation}
where $p_\mu$ is the 4-momentum of the photon at the disc/observer,
and $u^\mu$ is the 4-velocity of the disc/observer.  The photon
4-momentum is determined by the photon integration, and we suppose
that the observer is stationary (in Schwarzschild coordinates) and
very distant from the black hole; thus $u^\mu_{\rm obs}=(1,0,0,0)$.
Hence, we can compute $g$ for any given photon path provided we know
the 4-velocity field on the disc plane.  Initially, one might be
tempted to take the 3-velocity field directly from our MHD simulations
(which are approximately `Keplerian' with respect to the modified
potential) and construct the appropriate 4-velocity field as input for
our computation of $g$.  However, this approach is problematic for the
following reason.  Since our MHD simulation is intrinsically
non-relativistic, and based upon a potential that diverges at the event
horizon, the 3-velocities within the simulated plunging region can
significantly exceed those that would be found in a relativistic
calculation.  This would lead to artificially high values of $|g-1|$,
and hence artificially strong relativistic beaming.

Motivated by the fact that the velocity field in our simulated disc is
very close to Keplerian with respect to the pseudo-Newtonian
potential, we choose instead to use the correct relativistic 4-velocity for
test-particle circular motion in our computation of $g$.  This is well
defined down to the photon circular orbit at $3GM/c^2$, which lies
inside the inner edge of our computational grid.  Of course, disc material
within some radius close to the radius of marginal stability
(at $6GM/c^2$) will no longer be on circular orbits, but the deviations
in the relevant parts of the disc are small.  Employing this
assumption, we compute the function $g(x,y;i)$.

\section{Temporal variability of the simulated flux}

\subsection{The simulated lightcurve}
A direct calculation of the emission from the disc would require,
first, that we solve an energy equation, and second that we attain
sufficient resolution in a global simulation to model a stratified
disc and corona (Miller \& Stone 2000) where a significant fraction of
the dissipation probably occurs. Since our actual calculation is more
limited than this, we must instead choose some proxy for the
dissipation. Out of several possible proxies, we have opted to follow
Hawley \& Krolik (2001), and use the magnetic stress as a local tracer
of where in the flow dissipation is expected to occur
\footnote{Data storage limitations prompted us to make this basic 
choice {\em prior} to running full resolution simulations. 
During the runs, vertically integrated slices of $B_r B_\phi$ were the 
only quantities output at high time resolution.}. We then compute 
lightcurves and power spectra by making the further assumption 
that the dissipated energy is promptly radiated from the disc. 

The most important assumption is obviously the aforementioned one --
that {\em at a given radius, the azimuthal dependence of the emission
is proportional to the vertically integrated magnetic stress}. Even
given this assumption, however, there is some ambiguity as to the best
way to calculate the lightcurve. As described below, we calculate both
`raw' lightcurves, and lightcurves in which the average disc
dissipation has been normalized to match an analytic form.

\subsubsection{The raw lightcurve}

\begin{figure*}
 \psfig{figure=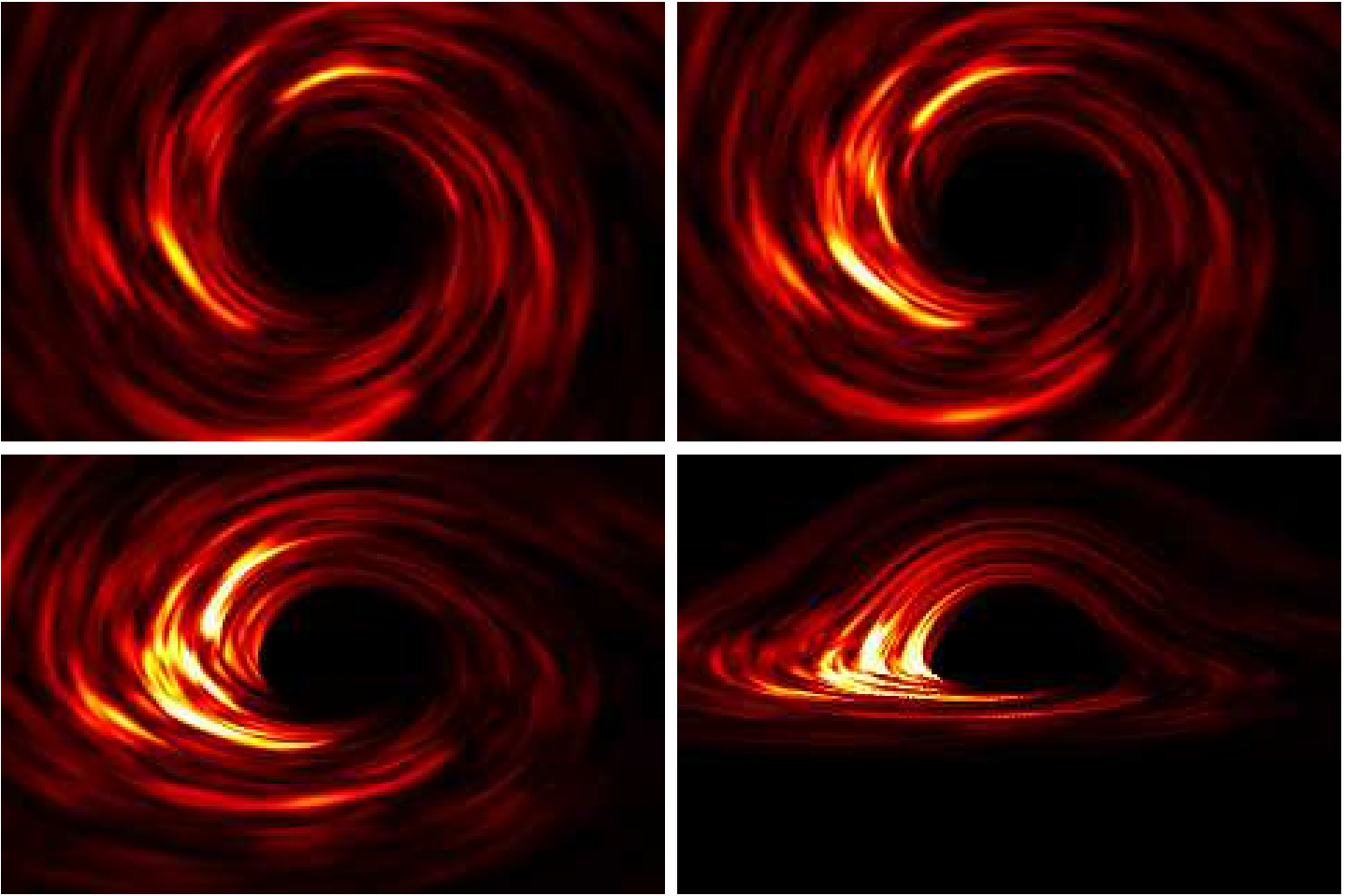,width=7.5truein,height=5.0truein}
 \caption{View of the disc as seen by a distant observer at 
          an inclination angle of 5$^\circ$ (upper left), 
          30$^\circ$ (upper right), 55$^\circ$ (lower left) 
          and 80$^\circ$ (lower right). In these raw images, 
          note the presence of stress extending to the inner 
          boundary of the computational domain, within 
          the marginally stable circular orbit. Movies 
	  showing the evolution of the simulated disc are 
	  available at {\tt http://jilawww.colorado.edu/$\sim$pja/black\_hole.html}.}
 \label{fig:fig_image}
\end{figure*} 

To compute the raw lightcurve, we first calculate the emission map 
$F_{\rm raw}$ that would be seen by observers orbiting with the disc 
flow. $F_{\rm raw}$ is given by (e.g. Hubeny \& Hubeny 1998).
\begin{equation} 
 F_{\rm raw} (r,\phi,t) = \sqrt{ {GM} \over r^3} \left( {A \over B} \right) 
 \int B_r B_\phi {\rm d}z,
\end{equation}
where the relativistic correction factors are,
\begin{eqnarray} 
 A & = & 1 - {{2 GM} \over {r c^2}} \nonumber \\
 B & = & 1 - {{3 GM} \over {r c^2}}.
\end{eqnarray}
Using the transfer function, which gives the mapping between
$(r,\phi)$ in the disc plane and $(x,y)$ in the image plane, we then
compute the image that would be seen by a distant observer viewing the
disc at a particular inclination angle, $i$. We assume that the
observer images the disc in a fixed spectral band, which introduces a
`K-correction' that depends upon the spectrum of the disc emission.
If at each radius the energy spectrum is $\propto \nu^{-1}$ (fairly
typically for accreting black holes), the resulting image is,
\begin{equation}
 I (x,y,t) \propto g^4 (x,y) F_{\rm raw} (r,\phi,t_{\rm emit}),
\end{equation} 
where one factor of $g$ comes from the K-correction.  We allow for the
varying flight time of photons from different parts of the accretion
disc by distinguishing between the time $t_{\rm emit}$, at which
photons are emitted from the disc, and time $t$ when the photons are
observed. Examples of these images are shown in
Figure~\ref{fig:fig_image}. The raw lightcurve at time $t_i$, ${\cal
F}(t_i)$ is then simply the sum over the image of the intensity at
time $t_i$ at each pixel $(x_j,y_k)$,
\begin{equation}
{\cal F}(t_i)=\sum_{j,k}I(x_j,y_k,t_i).
\end{equation}
We emphasize that the predicted variability properties depend 
upon the assumed emission spectrum from the disc as well as on the 
statistical fluctuations of MHD disc turbulence. In particular, 
modeling blackbody emission from the disc would require 
a different radial weighting of the simulated disc 
dissipation.

\begin{figure*}
\hbox{
 \psfig{figure=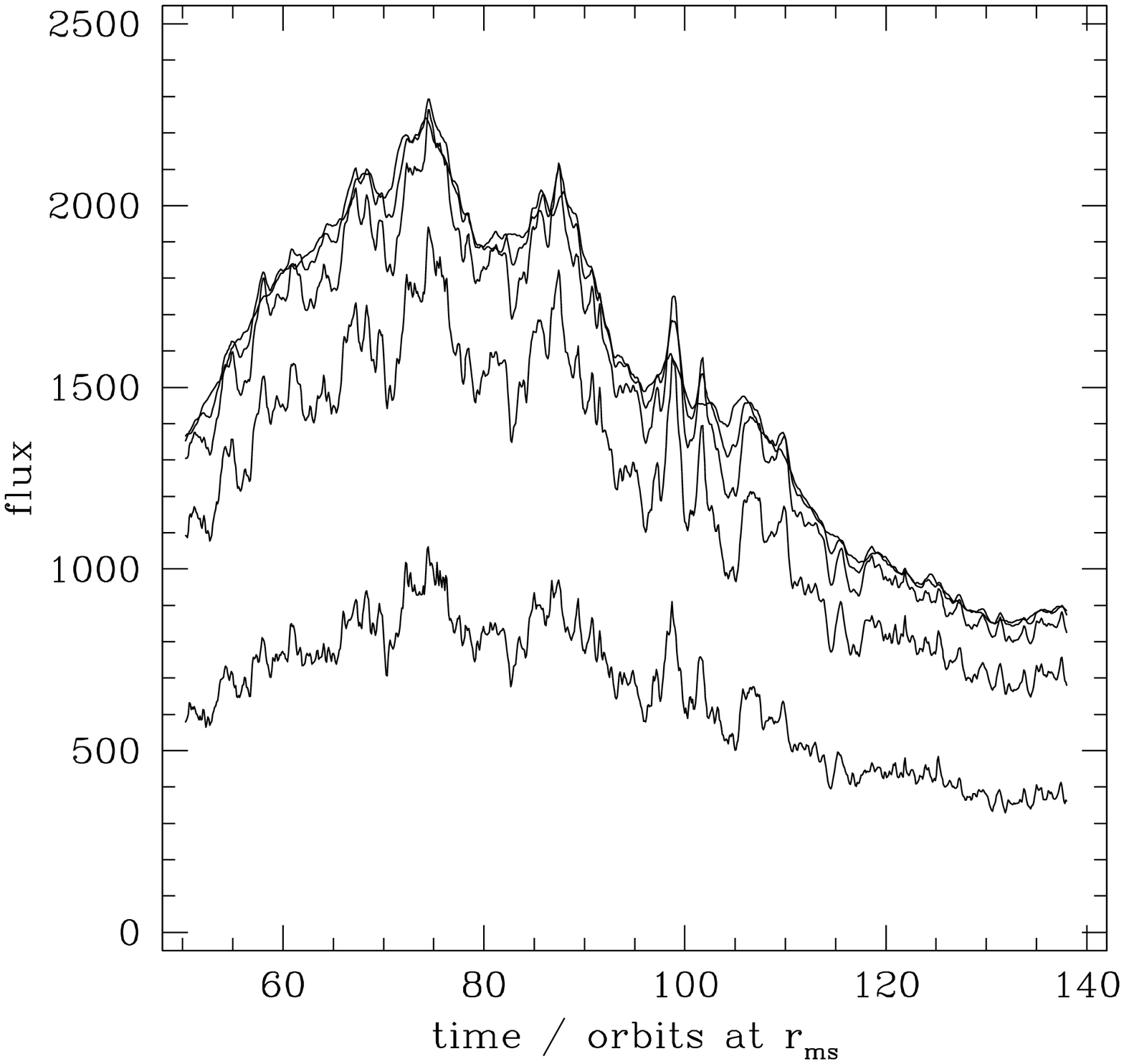,width=3.5truein,height=3.5truein}
 \psfig{figure=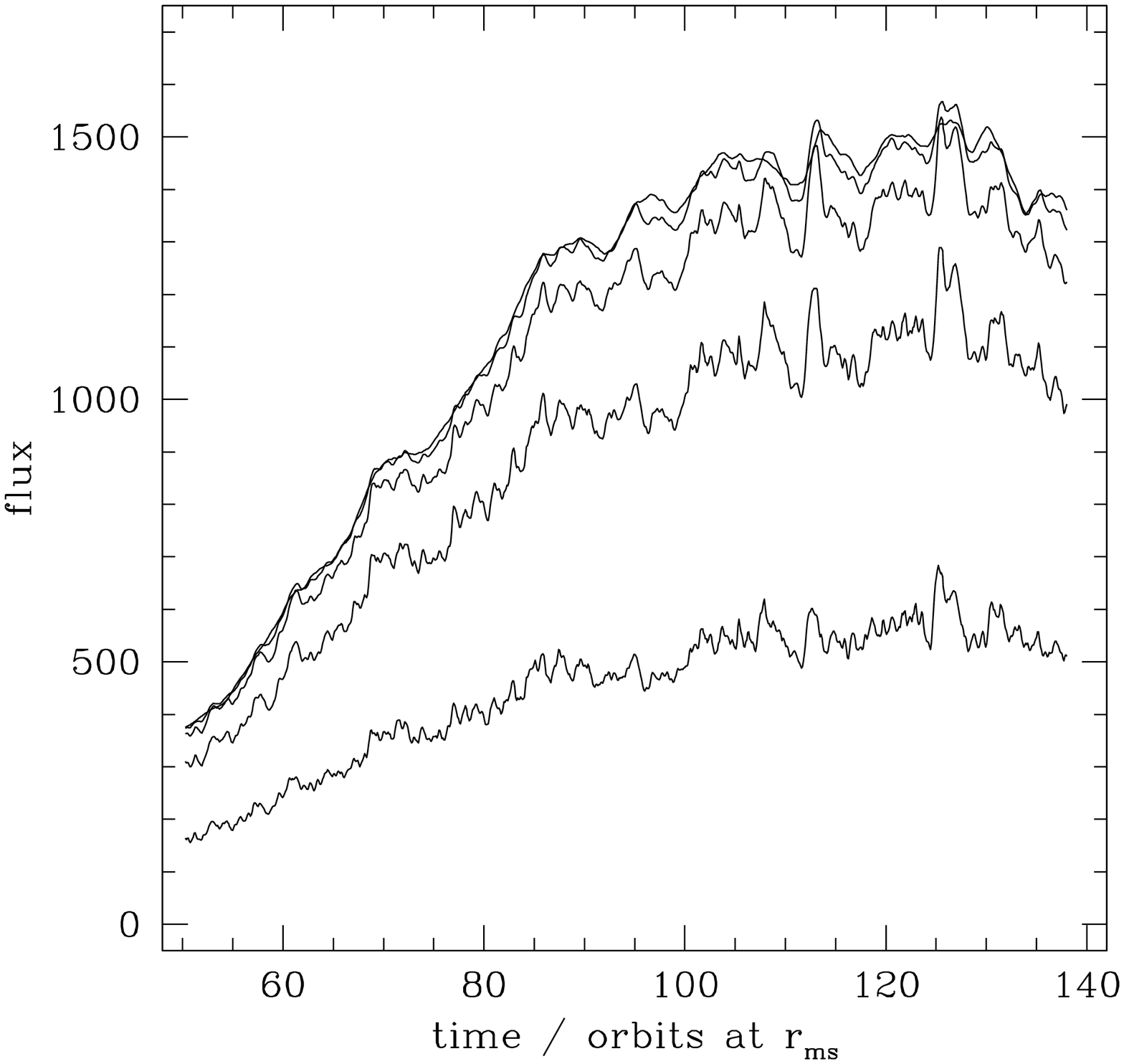,width=3.5truein,height=3.5truein}
}
 \caption{Raw lightcurves showing the
   integrated flux from the disc as a function of the system
   inclination. From top downward, $i=1^\circ, 20^\circ, 40^\circ,
   60^\circ, 80^\circ$. The small scale fluctuations increase with
   increasing system inclination. The hot simulation is displayed in
   the left panel, whereas the right panel displays the cold
   simulation.}  
\label{fig:fig_flux}
\end{figure*}

The resulting light curves are shown in Figure~\ref{fig:fig_flux} for
several different inclinations. As already noted, the shape of 
these lightcurves over the longest timescales reflects only 
the viscous evolution of the ring of gas used for our initial 
conditions. There is a secular increase in the inner accretion 
rate (and accompanying dissipation) as the MRI saturates at 
increasingly large radii, followed (for the hot simulation) 
by an eventual decline once a substantial fraction of the 
gas has been accreted. These slow changes in accretion rate are 
artefacts of the initial conditions used in the simulation --- real
accretion discs are long lived objects with very large reservoirs of
material. Of more relevance to real systems is the inclination
dependence of these light curves. One observes a marked increase in
rapid variability for the more edge on systems. This is readily
interpreted as the effects of relativistic beaming coupled with the
`blobby' nature of the simulated emission.  We shall explore this
point in more detail below when we discuss temporal power spectra.

\subsubsection{Novikov-Thorne normalized lightcurve}
The raw lightcurve computed from the simulation self-consistently 
includes the contribution from torque at the marginally stable orbit 
to the emission of the inner disc. As shown in Figure~\ref{fig_torque}, 
this contribution can be important, especially for hotter (and 
therefore geometrically thicker) discs (Reynolds \& Armitage 2001). However, 
because the domain of the simulation is limited, only the inner 
region of the simulated disc is in a steady-state. The raw 
lightcurve therefore {\em underestimates} the contribution 
to the flux from larger radii in the disc.

Cognisant of this drawback, we also compute lightcurves by combining 
the simulated fluctuations in the stress (in $\phi$ and $t$) with an 
analytic dissipation profile due to Novikov \& Thorne (1973). Specializing 
to the case of a Schwarzschild black hole, the radial dependence of the 
disc flux measured by an observer orbiting with the disc gas is (Page \& 
Thorne 1974),
\begin{eqnarray}
 F_{\rm NT} (r) \propto {1 \over { x^4 (x^3 - 3 x)} } 
 \left[ x - \sqrt{6} - { \sqrt{3} \over 2 } \ln 
 \left( { {x - \sqrt{3}} \over {\sqrt{6} - \sqrt{3}} } \right) \right. \nonumber \\
 + \left. {\sqrt{3} \over 2} \ln \left( { {x + \sqrt{3}} \over {\sqrt{6} + \sqrt{3}} }, 
 \right) \right]
\end{eqnarray} 
where,
\begin{equation} 
 x \equiv \sqrt{ { r \over {(GM / c^2)} } }.
\end{equation} 
Given this dissipation profile, we proceed to construct a lightcurve as 
before, except that we replace the emission map $F_{\rm raw}(r,\phi,t)$ 
with $F_{\rm NT}(r,\phi,t)$, defined as,
\begin{equation}
 F_{\rm NT}(r,\phi,t) = { {B_r B_\phi (r,\phi,t)} \over {<B_r B_\phi> (r)} } F_{\rm NT} (r).
\end{equation} 
In this expression $B_r B_\phi$ in both the numerator and the
denominator is understood to be vertically integrated, and the angle
brackets denote averaging over both $\phi$ and $t$. By construction,
the resulting dissipation map $F_{\rm NT}(r,\phi,t)$ has the same
time-averaged dissipation profile as a Novikov-Thorne disc, on top of
which are superimposed fluctuations whose statistical properties are 
computed directly from the MHD simulation.  Using this normalized 
dissipation map (which we shall
refer to as a `NT-normalization'), one can construct the observed
emission map and a lightcurve in exactly the manner discussed above.

\subsection{The temporal power spectrum} 

\begin{figure}
\hbox{
 \psfig{figure=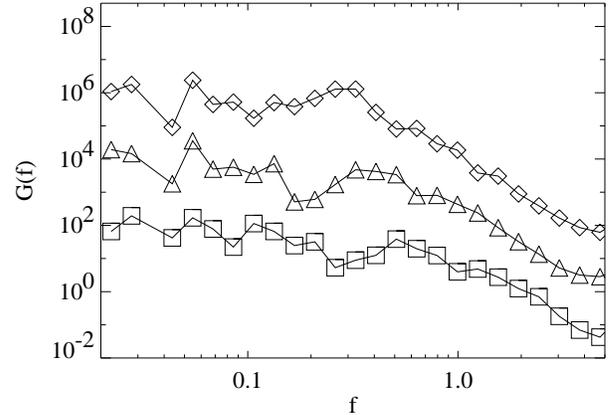,width=0.5\textwidth}
}
\caption{Power spectra of the predicted emission from individual annuli 
  of the hot disc simulation. The system inclination is $i=50^\circ$.
  From bottom upward, we show results for annuli defined by $6 < r /
  r_{\rm ms} < 7$, $8 < r / r_{\rm ms} < 9$, and $10 < r / r_{\rm ms}
  < 11$. Below a break frequency, which is comparable to the orbital
  frequency, the power is flat or slowly declining (roughly bracketed
  by $G(f) = {\rm const}$ and $G(f) \propto \nu^{-1}$).  At high
  frequencies, $G(f) \propto \nu^{-3.5}$.}
\label{fig:fig_annulus}
\end{figure} 

\begin{figure*}
\hbox{
\psfig{figure=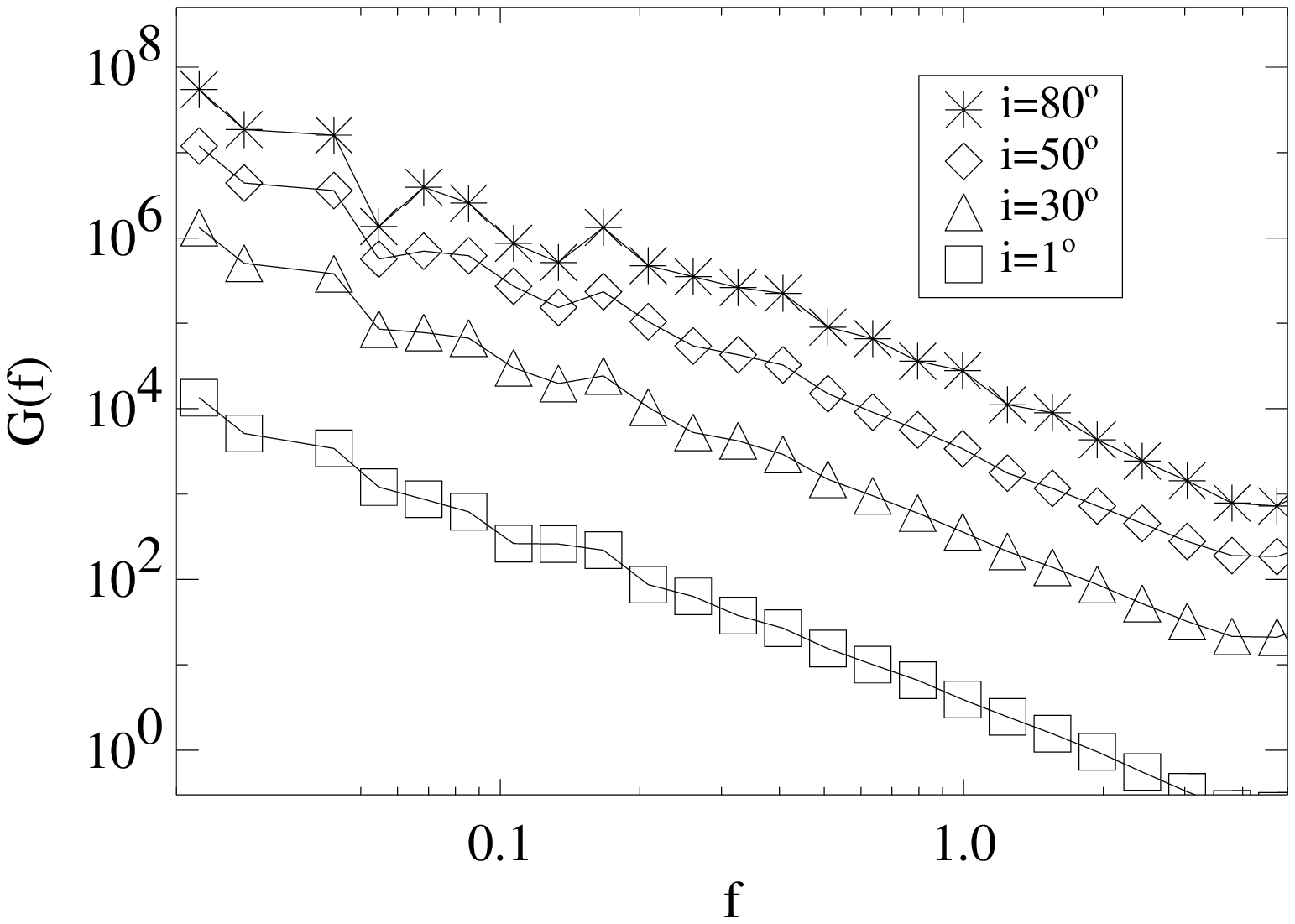,width=0.5\textwidth}
\psfig{figure=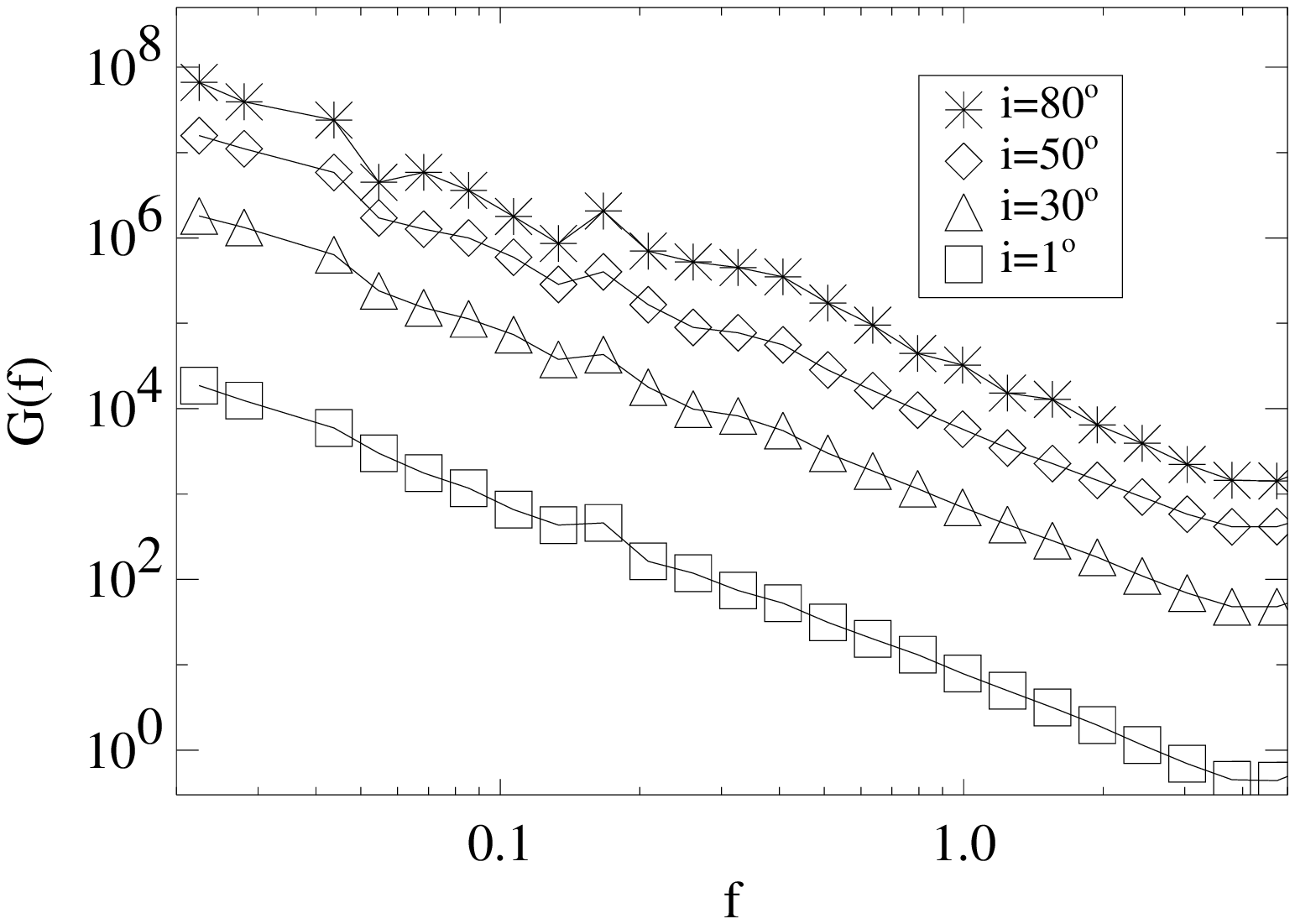,width=0.5\textwidth}
}
\hbox{
\psfig{figure=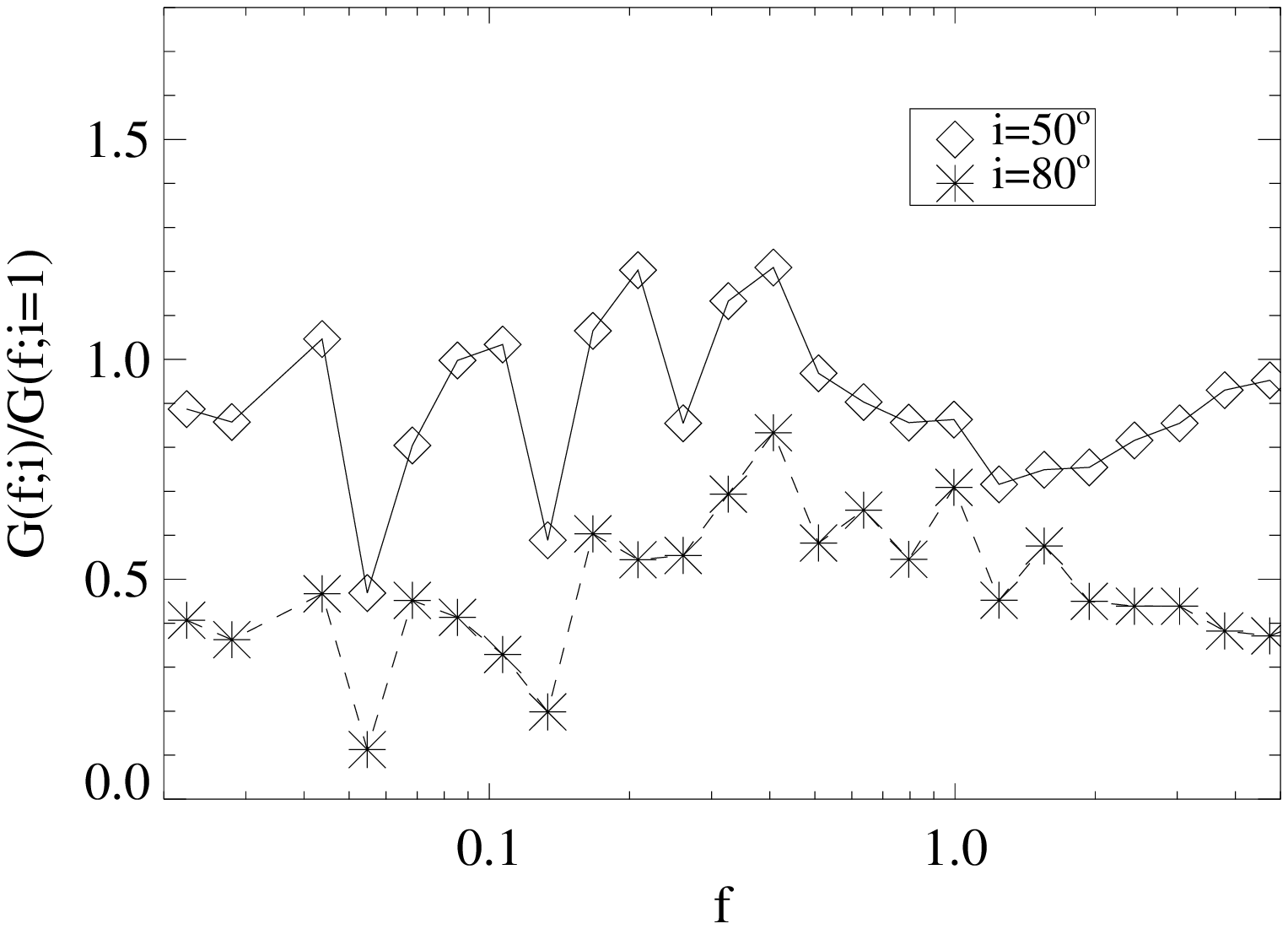,width=0.5\textwidth}
\psfig{figure=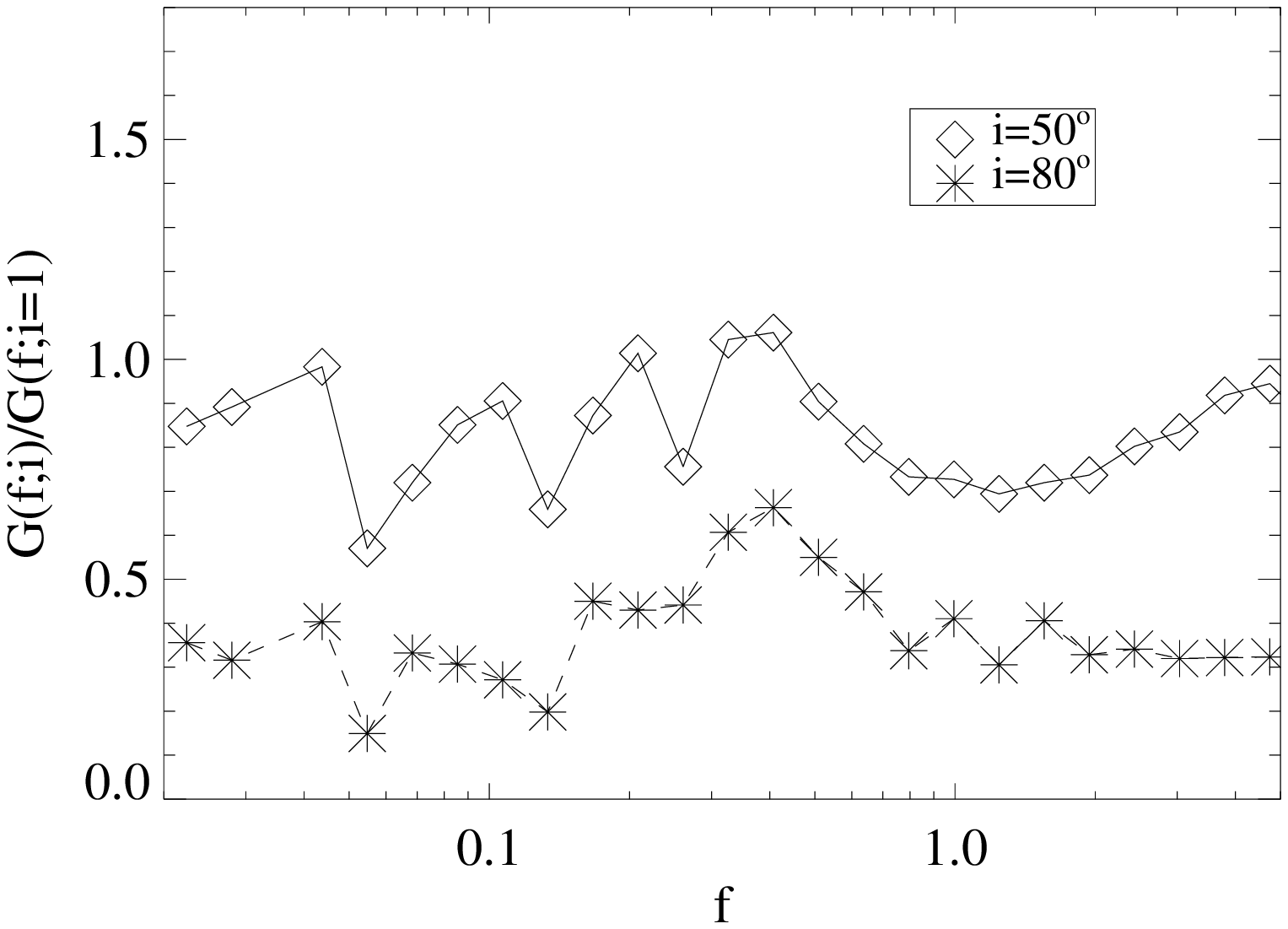,width=0.5\textwidth}
}
\caption{{\it Upper panels:} Power spectra $G(f)$ for the raw (left) 
  and NT-normalized (right) light curves resulting from the cold
  simulation.  The frequency $f$ has been rescaled such that the
  orbital frequency at the radius of marginal stability corresponds to
  $f=1$.  Shown here are the power spectra for $i=1^\circ$,
  $i=30^\circ$ (offset by $10^2$), $i=50^\circ$ (offset by $10^3$),
  and $i=80^\circ$ (offset by $10^4$). {\it Lower panels:} The 
  ratio $G(f,i)/G(f,i=1)$ for $i=50^\circ$ and $i=80^\circ$.
  This representation helps isolate the effect of inclination on the
  power spectrum. See text for discussion.}
\label{fig:powerspec_cold}
\end{figure*}

\begin{figure*}
\hbox{
\psfig{figure=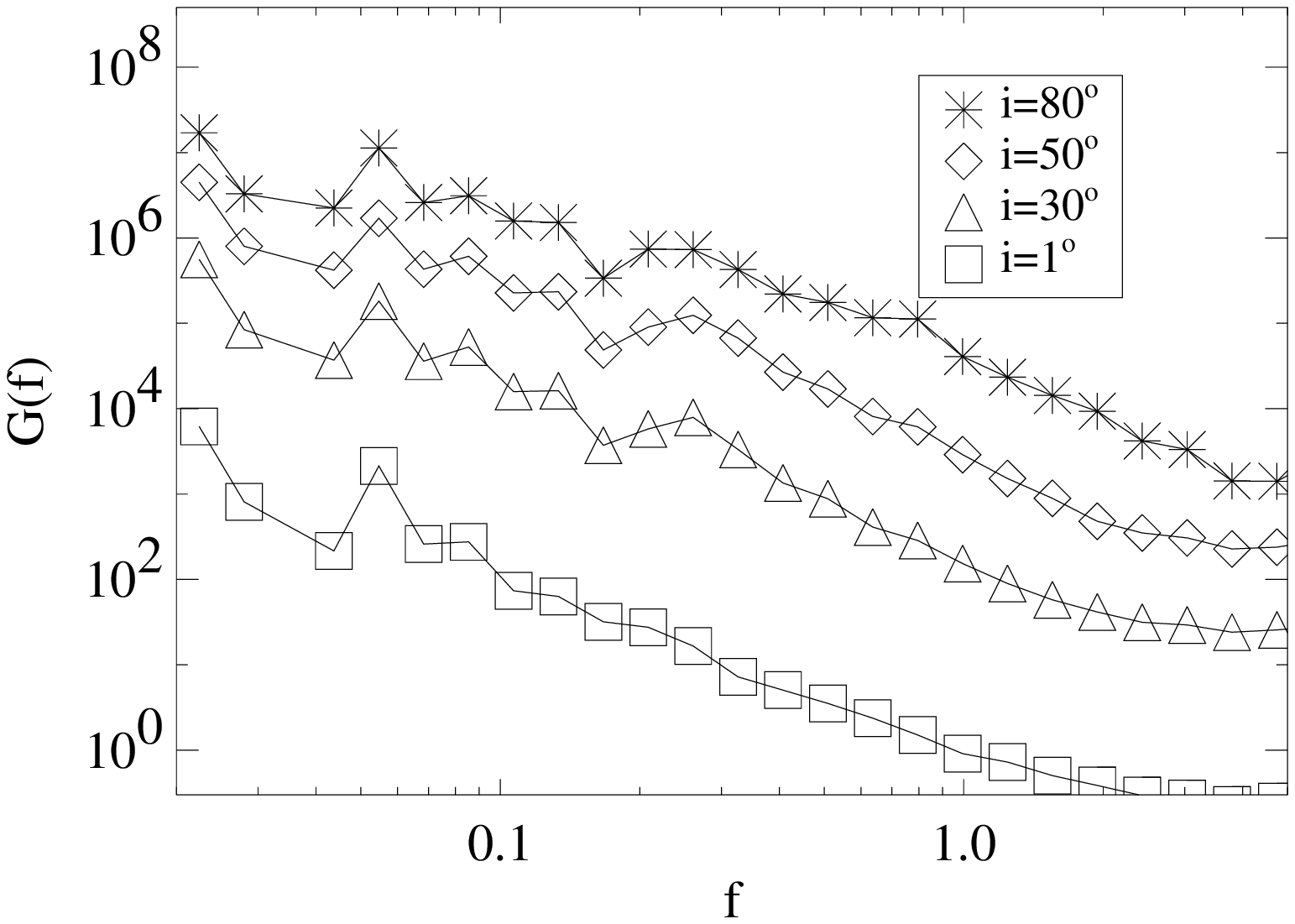,width=0.5\textwidth}
\psfig{figure=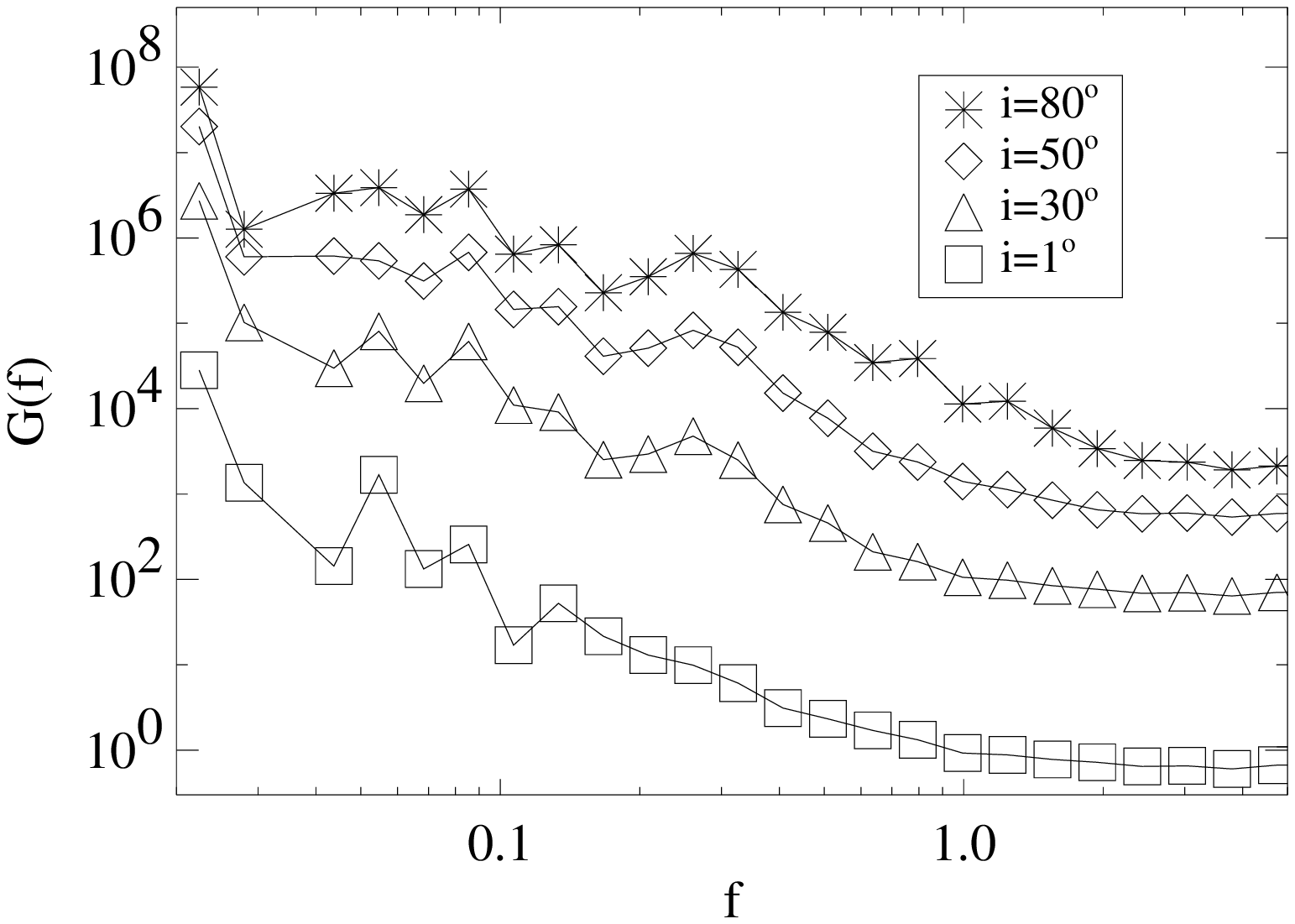,width=0.5\textwidth}
}
\caption{Same as the upper panels in Fig.~\ref{fig:powerspec_cold}, 
except here computed for the hot simulation.}
\label{fig:powerspec_hot}
\end{figure*}

For a given lightcurve, we compute the temporal power spectrum $G(f)$, 
where $f$ is the frequency, via a straightforward Fourier
transform together with the ``Leahy normalization'',
\begin{equation}
G(f)=\frac{2|\sum_i {\cal F}(t_i)\exp(2\pi j f t_i)|^2}{\sum_i
{\cal F}(t_i)}.
\end{equation}
We truncate the resulting power spectra at the
high-frequency end at the Nyquist frequency (i.e., half of the
sampling frequency of the lightcurve) to account for the 
finite sampling time.  In addition, frequencies are rescaled 
such that the orbital frequency at the radius of
marginal stability is unity.

Figure \ref{fig:fig_annulus} shows power spectra computed using 
emission that originates from several narrow annuli in the inner 
disc of the hot simulation. The power spectra are described 
well by broken power laws. Below a break frequency, which 
is comparable to the local orbital frequency, $G(f)$ is 
flat or modestly declining ($G(f) \propto f^{-1}$). At 
higher frequencies, above the break, we obtain a steep 
spectrum with $G(f) \propto f^{-3.5}$.

Power spectra computed using the integrated disc emission (ie directly
from the lightcurves in Figure~\ref{fig:fig_flux}) are shown in
Figures~\ref{fig:powerspec_cold} and \ref{fig:powerspec_hot} for the
cold and hot runs respectively. For the cold run, $G(f)$ calculated
from the raw lightcurve is accurately approximated by a power law with
an index of -2 for all frequencies which are accessible using the
simulation.  There are no significant changes in the slope of $G(f)$
with increasing system inclination.  Figure \ref{fig:powerspec_cold}
also shows power spectra computed from the NT-normalized lightcurves.
Although not identical, these are almost indistinguishable from power
spectra calculated using the raw lightcurve. We conclude that the
presence of torque at the marginally stable orbit -- at least at the
rather modest levels obtained in our calculations -- does not
significantly affect the predicted power spectrum.

The power spectra shown in Figure \ref{fig:powerspec_cold} appear to
be accurately described by single power laws over the range of
frequencies accessible using the simulation. Observationally, high
frequency QPOs in black hole systems are typically weak features.
For example, the broad QPO at 300~Hz reported by 
Remillard et al. (1999) in GRO~J1655-40 had an amplitude of 
less that 1\%. Substantially longer simulations would be 
needed before we could measure the theoretical power 
spectrum accurately enough to detect broad QPOs of such small 
amplitude. We can, however, place
limits on the possible existence of stronger QPOs that could arise 
from the `lighthouse' mechanism discussed by Abramowicz et al. 
(1991) and Karas (1999). Such QPOs, since they originate
from the combination of hotspots with relativistic beaming, should
become stronger at higher system inclinations. To investigate this
possibility, the lower panels of Figure \ref{fig:powerspec_cold} show
the relative power for inclined systems as compared to face-on
systems. For $i=50^\circ$, any QPOs at frequencies comparable to the
orbital frequency contribute at most $\approx 20$\% of the broadband
power.

Figure \ref{fig:powerspec_hot} shows the corresponding power spectra
extracted from the hot simulation. The results are generally
consistent, though the deviations from a power law form for $G(f)$ are
larger. At higher inclinations there are some indications of the
presence of excess power at a frequency of around 0.2--0.3 times the
orbital frequency at $r_{\rm ms}$. Excess power is also seen in the 
cold simulation at similar frequencies, though at a lower level.  
Longer simulations are required to investigate the 
reality of such weak, broad features.

Since the emission from a particular annulus has a
power-spectrum that breaks at the orbital frequency of that annulus,
we expect that the disk-integrated power-spectrum should possess a
high-frequency break corresponding to the innermost emitting annulus.
The overall power-spectra should therefore steepen
appreciably above $f\sim 1$. In fact, we do not see such a
high-frequency break in Figures~\ref{fig:powerspec_cold} and
\ref{fig:powerspec_hot}. We have investigated the source of 
this additional high frequency noise in our lightcurves, and 
find that it originates from emission from the outermost 
parts of the simulated disc. Our suspicion is that the 
spatial discretization of the disc into cells -- which 
are physically largest in the outer disc -- introduces 
a low level of high frequency flicker into the lightcurve.
This hampers our ability to see high-frequency cutoffs 
in the disk-integrated power-spectra.

\section{Iron line variability}

The broad iron line is one of the best understood observational probes
of black hole accretion discs at the present time (e.g., see reviews
by Fabian et al. 2000; Reynolds \& Nowak 2003).  Thus, it is useful to
examine the predictions of our simulation for temporal variability of
the iron line.

\begin{figure*}
\hbox{
\psfig{figure=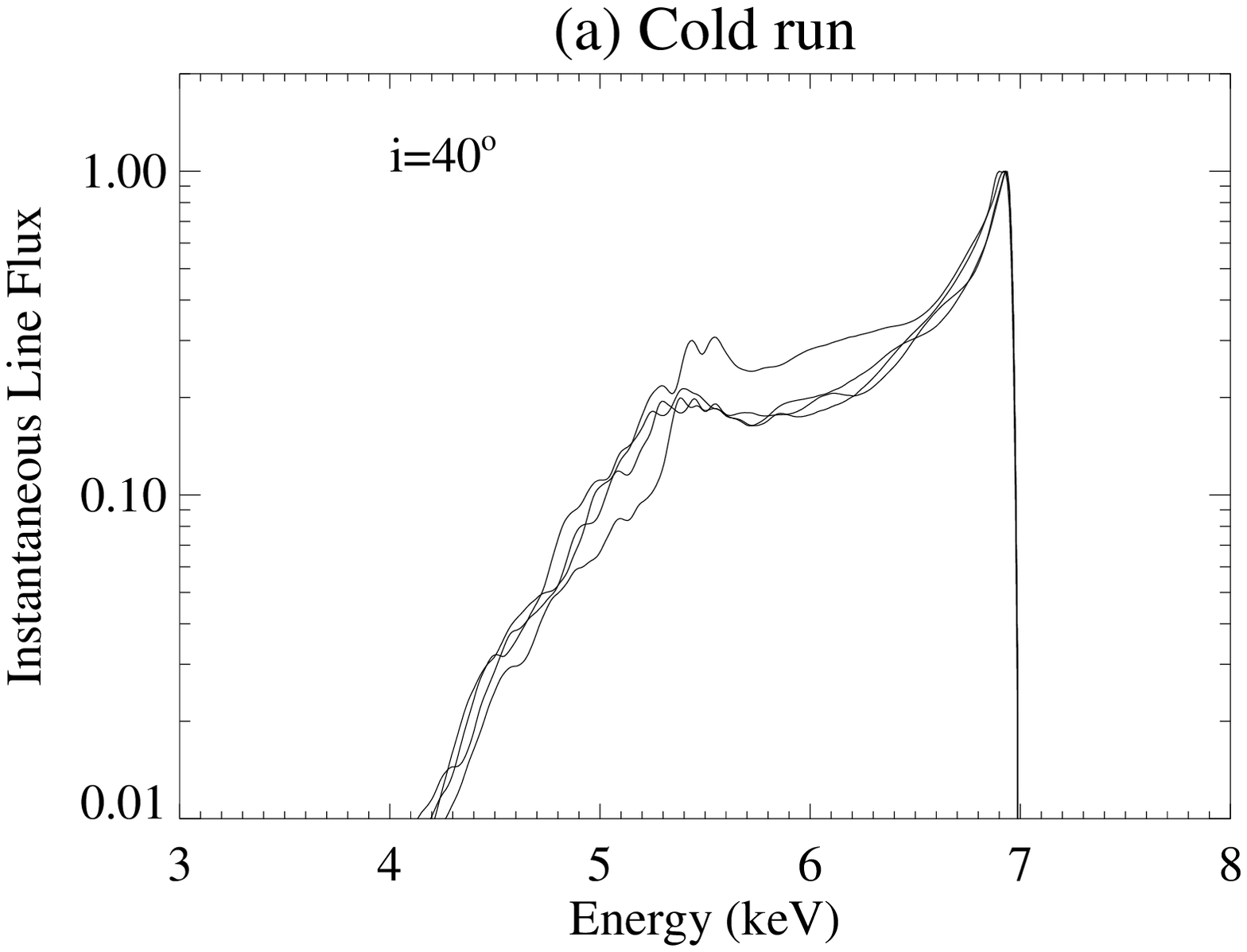,width=0.5\textwidth}
\psfig{figure=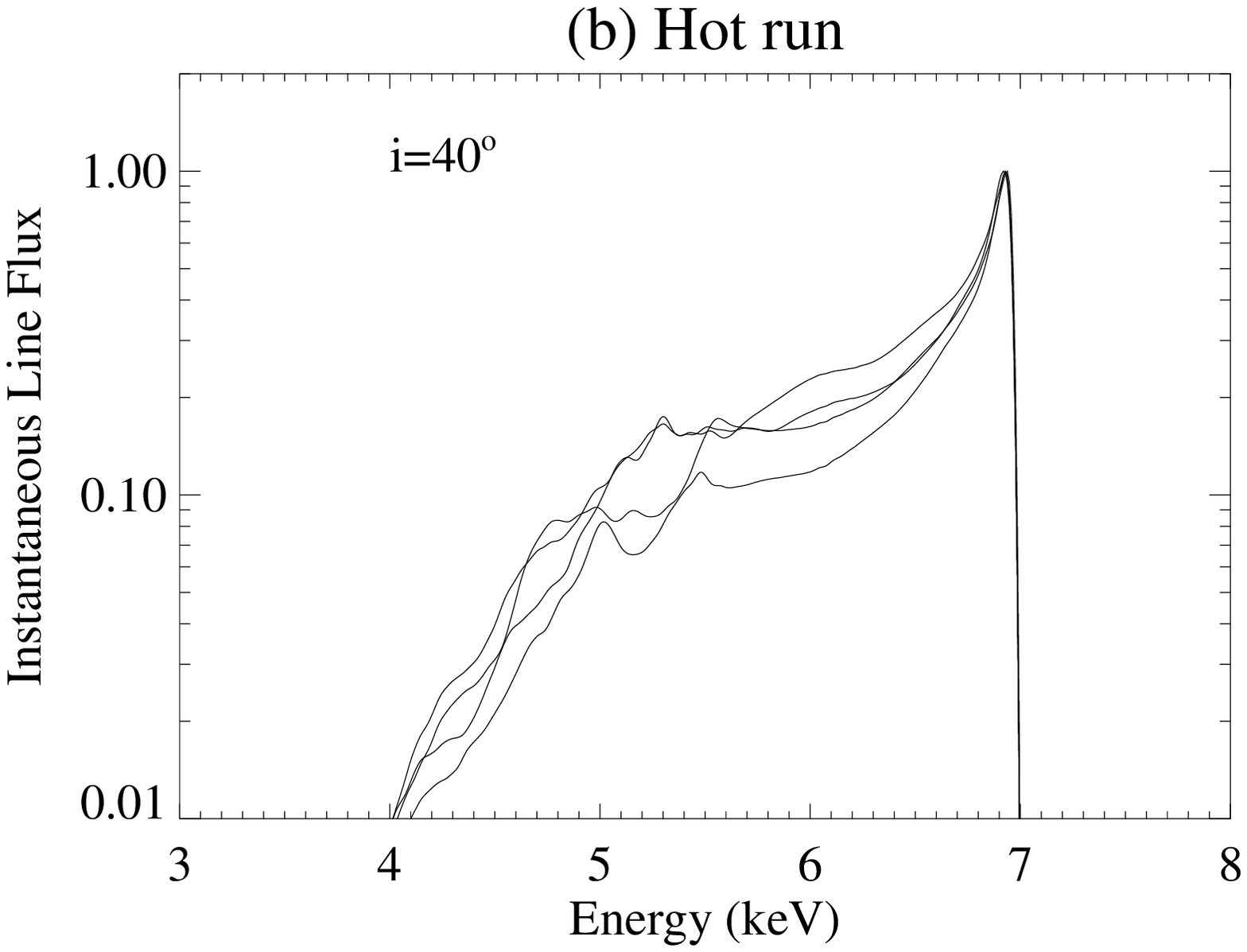,width=0.5\textwidth}
}
\caption{Instantaneous simulated iron line profiles for the cold (left) 
and hot (right) simulations. Note that the line flux is plotted on 
a logarithmic vertical scale.}
\label{fig:ironline}
\end{figure*}

\begin{figure}
  \hbox{\psfig{figure=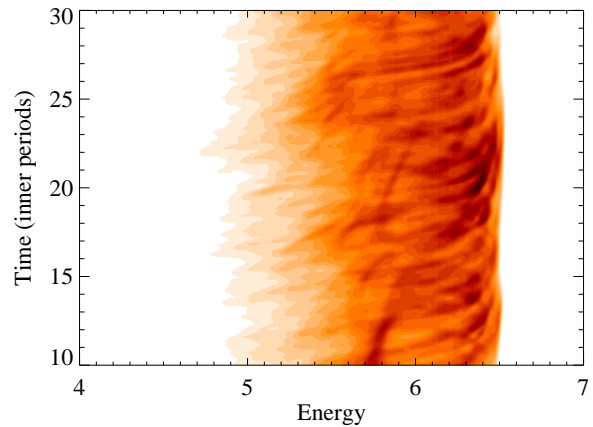,width=0.5\textwidth} }
\caption{Iron line intensity as a function of energy and time for a 
20 orbit segment of our simulation, assuming an inclination of
$i=20^\circ$.}
\label{fig:linevar}
\end{figure}

The predicted temporal variability of the iron line obviously 
depends on how that line is excited. Different authors have 
computed the iron line response to flares in the inner 
regions of the flow, using a variety of different geometries 
and kinematics for the source of exciting photons (Fabian et al. 1989; 
Stella 1990, Matt \& Perola 1992; Reynolds et al. 1999; Ruszkowski 2000).
In this work, we assume that the local iron line intensity coming 
from the surface of the disc is proportional to the local continuum 
intensity (and, hence, is proportional to the local vertically 
integrated azimuthal magnetic stress).  The line intensity at 
image-plane position $(x,y)$, time $t$ and energy $E$ is then given by
\begin{equation}
I_{\rm line}(x,y,t;E)\propto g^4 (x,y) F(r,\phi,t_{\rm emit}) \delta(E-gE_0),
\end{equation}
where $E_0$ is the rest-frame energy of the emission line and
$F(r,\phi,t)$ is taken to be either the raw of NT-normalized disc
flux.  Note that the K-correction is not relevant for computing line
fluxes and, in this expression, one power of $g$ results from the
transformation of the frequency-element.  The total simulated observed
line is obtained from integrating this expression over the image
plane.  Figure~\ref{fig:ironline} shows instantaneous iron line
profile resulting from the hot simulation for an inclination of
$i=40^\circ$.  The skewed and redshifted profile is caused by a
combination of the line-of-sight Doppler effect, the transverse Dopper
effect (i.e. special relativistic time dilation) and gravitational
redshifts (see Fabian et al. 2000 and Reynolds \& Nowak 2003 for a
detailed discussion of the physics determining these line profiles).   

Under the assumptions employed above, the simulated iron line at any
particular instant is a weighted redshift map of the turbulent
emissivity pattern across the disc.  Thus, the temporal and azimuthal
variability caused by the MHD turbulence will produce a level of
inevitable and irreducible temporal variability in the iron line
profile.  This can be seen by comparing the instantaneous line
profiles in Figure~\ref{fig:ironline}. From this figure, it can be seen
that the iron line from the hot disc displays dramatic (factor of
$\sim 2$) variability between independent time slices, whereas the
line profile from the cold disc is more stable, showing deviations at
only the $\sim 30$ percent level.  

The nature of the iron line variability is demonstrated more clearly
in Figure~\ref{fig:linevar}, which shows the intensity of the
simulated iron line emission as a function of energy and time for a
representative interval during our simulations, assuming an
inclination of $i=20^\circ$. This is a plausible inclination for many
Seyfert-1 nuclei, the systems in which broad iron lines have been best
studied to date.  One can see a tremendous degree of structure in the
iron line profile as a function of both energy and time. Of
particular note are the distinctive arcs that correspond to the orbital
motion of especially bright regions within the turbulent flow, visible 
in Figure~\ref{fig:fig_image}. These arcs essentially trace out (in the 
energy-time plane) the paths of test particle with orbits close 
to the radius of marginally stability. If they could be detected, 
they would provide a quantitative test of the flow extremely 
close to the black hole. High temporal resolution is 
required to observe such features, but the required capabilities 
should be readily within reach of the planned {\it Constellation-X} 
and {\it XEUS} missions.

In fact, observational evidence for transient non-axisymmetric
structure already exists.  {\it Chandra}-HETG observations of the
Seyfert-1 galaxy NGC~3516 have found substructure within the well
known broad iron line profile of this object (Turner et al. 2002).  
The substructure has the form of narrow line-line features at
rest-frame energies of 5.6\,keV, 6.2\,keV and 7.1\,keV.  Plausible
explanations exist for the latter two features; the 7.1\,keV feature
can be readily attributed to the fluorescent K$\beta$ emission of cold
iron, and the 6.2\,keV feature might be the `Compton shoulder' of
the prominent 6.4\,keV K$\alpha$ fluorescent line of cold iron.
However, the 5.6\,keV feature cannot be attributed to any expected
features.  Turner et al. suggest that this feature is, in fact, iron
K$\alpha$ fluorescence with a rest-frame energy of 6.4\,keV from an
anomalously bright annulus or spot on the accretion disc surface.  The
flux contained in the 5.6\,keV feature is approximately 5\% of that
contained in the overall broad iron line.

Is it possible to relate the substructure observed in the NGC~3516
iron line with the turbulence induced substructure seen in our
simulations?  While large amplitude substructure is present in
virtually every {\it instantaneous} simulated line profile, it is much
diminished once the line profiles are averaged over more than a few
orbits.  The {\it Chandra}-HETG observation of Turner et al. that
yielded the clear detection of the iron line substructure was $8\times
10^4$\,s in duration. Assuming a black hole mass of $2.3\times
10^7\,{\rm M}_\odot$ (estimated using the velocity dispersion 
from di Nella et al. 1995, in conjunction with the scaling relations of 
Gebhardt et al. 2000 or Ferrarese \& Merritt 2000)\footnote{Independent  
estimates of the black hole mass are comparable. For example, Onken et al. (2003) 
estimate $M_{BH} = 1.7 \times 10^7 \ M_\odot$ from reverberation 
mapping.}, 
this corresponds to approximately 20 orbital periods of the accretion disc 
at the radius of marginal stability.  Thus, to compare our simulations to these
data, we averaged our simulated line profiles over 20 orbits of the
inner disc.  We assumed a disc inclination of $i=40^\circ$ (Wu \& Han
2001).  Given the useful length of our simulation, five independent
20-orbit line profiles could be formed.  We found that 4 out of 5 of
these profiles possessed no strong deviations from the average.  In
one profile, however, there was a slight excess centered at an energy
of 5.5\,keV, possessing 2--3\% of the total flux.  Therefore, while
the true interpretation of the {\it Chandra} data for NGC~3516 remains
unclear, turbulence-induced substructure is a viable possibility even
when one considers that the disc has likely undergone several orbits
during the observation.

\section{Summary}

In this paper, we have explored the predicted variability of 
accretion discs around Schwarzschild black holes. The 
temporal and spatial fluctuations in the magnetic stress which 
drives accretion have been calculated using three-dimensional global 
simulations in a pseudo-Newtonian potential, which ought to 
provide a reasonable first approximation to the disc dynamics 
outside the marginally stable orbit. By making the further 
critical assumption that the radiation from the disc follows 
the local stress, we have calculated the predicted emission that 
would be seen by an observer using a fully relativistic ray-tracing 
code that accounts for the relativistic effects of beaming, 
gravitational redshift, and light bending.

Our results for the temporal power spectrum of the predicted 
emission are consistent with those of previous authors 
(Kawaguchi et al. 2000; Hawley \& Krolik 2001, 2002). Fluctuations 
in the magnetic stress, when integrated across the disc, produce a high 
frequency power spectrum described by a power law with slope $f^{-2}$. 
To a first approximation, observers viewing the system at a wide range of 
inclinations -- from face-on to almost edge-on -- are predicted to derive 
similar power spectra. In more detail, high inclinations are found 
to boost the relative power at frequencies a factor of a few 
below the orbital frequency at the marginally stable orbit.
There is little change in the predicted 
power spectra due to the presence or absence of modest levels 
of magnetic stress at the marginally stable orbit. No strong 
QPOs are seen in the simulations, though weak features could 
be present and would not be clearly observed given the limited statistics 
available from $\approx 10^2$ orbit simulations. 

The broad iron line provides an alternate probe of conditions 
in the inner disc, and has the advantage of retaining velocity 
information. We have used the simulations to quantify the 
predicted iron line variability within a model in which the 
local iron line emissivity tracks the locally generated 
continuum emission. This model represents the opposite 
extreme from reverberation models, in which the source of line 
exciting flux is often assumed to be widely spatially separated 
from the disc. We find that order unity fluctuations in the 
line flux at fixed energy occur for discs with $h/r \sim 0.1$, 
with the fluctuation amplitude increasing rapidly with 
increasing disc thickness. Averaging over {\em at least} several 
orbits at the radii of interest is needed before meaningful 
averages can be extracted (see also Papaloizou \& Nelson 2003; Winters, 
Balbus \& Hawley 2003). 
Observationally, this implies that long integrations, extending 
over tens of orbital timescales, are required in order 
to reduce the fluctuation amplitude to percent levels and 
thereby recover the mean line profile.

\section*{Acknowledgements}
We thank Cole Miller, Eve Ostriker, Daniel Proga and Mateusz
Ruszkowski for extensive discussions, and Pete Ruprecht for assisting
with computational support. We thank the referee for providing 
an extremely prompt report on the paper. CSR acknowledges support from the National
Science Foundation under grant AST0205990.


\begin{thebibliography}{}

\bibitem{}
Abramowicz M.A., Bao G., Lanza A., Zhang X.-H., 1991, A\&A, 245, 454

\bibitem{}
Afshordi N., Paczynski B., 2003, ApJ, submitted, astro-ph/0202409 

\bibitem{}
Agol E., Krolik J.H., 2000, ApJ, 528, 161

\bibitem{}
Arlt R., Rudiger G., 2001, A\&A, 374, 1035

\bibitem{}
Armitage P.J., 1998, ApJ, 501, L189

\bibitem{}
Armitage P.J., Reynolds C.S., Chiang J., 2001, ApJ, 548, 868

\bibitem{}
Balbus S.A., Hawley J.F., 1991, ApJ, 376, 214

\bibitem{}
Balbus S.A., Hawley J.F., 1998, Rev. Mod. Phys., 70, 1

\bibitem{}
Balucinska-Church M., Church M.J., 2000, MNRAS, 312, L55

\bibitem{}
Brandenburg A., Nordlund A., Stein R.F., Torkelsson U., 1995, ApJ, 446, 741

\bibitem{}
di Nella, H.; Garcia, A. M.; Garnier, R.; Paturel, G., 1995, A\&AS, 113, 151

\bibitem{}
Evans C.R., Hawley J.F., 1988, ApJ, 332, 659

\bibitem{}
Fabian A.C., Rees M.J., Stella L., White N.E., 1989, MNRAS, 238, 729

\bibitem{}
Fabian A.C., Iwasawa K., Reynolds C.S., Young A.J., 2000, PASP, 112, 1145

\bibitem{}
Ferrarese L., Merritt D., 2000, ApJ, 539, L9

\bibitem{}
Gammie C.F., 1999, ApJ, 522, L57

\bibitem{}
Gebhardt K. et al., 2000, ApJ, 539, L13

\bibitem{}
Hawley J.F., 2000, ApJ, 528, 462

\bibitem{}
Hawley J.F., 2001, ApJ, 554, 534

\bibitem{}
Hawley J.F., Balbus S.A., Stone J.M., 2001, ApJ, 554, L49

\bibitem{}
Hawley J.F., Gammie C.F., Balbus S.A., 1995, ApJ, 440, 742

\bibitem{}
Hawley J.F., Krolik J.H., 2001, ApJ, 548, 348

\bibitem{}
Hawley J.F., Krolik J.H., 2002, ApJ, 566, 164

\bibitem{}
Hubeny I., Hubeny V., 1998, ApJ, 505, 558

\bibitem{}
Igumenshchev I.V., Narayan R., 2002, ApJ, 566, 137

\bibitem{}
Igumenshchev I.V., Narayan R., Abramowicz M.A., 2003, ApJ, submitted, astro-ph/0301402

\bibitem{}
Iwasawa K., 1996, MNRAS, 282, 1038

\bibitem{}
Karas V., 1999, PASJ, 51, 317

\bibitem{}
Kawaguchi T., Mineshige S., Machida M., Matsumoto R., Shibata K., 2000, PASJ, 52, L1

\bibitem{}
Krolik J.H., 1999, ApJ, 515, L73

\bibitem{}
Lee J.C., Iwasawa K., Houck J.C., Fabian A.C., Marshall H.L., Canizares C.R., 
2002, ApJ, 570, L47

\bibitem{}
Lubow S.H., Papaloizou J.C.B., Pringle J.E., 1994, MNRAS, 267, 235

\bibitem{}
Machida M., Matsumoto R., 2003, ApJ, in press, astro-ph/0211240

\bibitem{}
Machida M., Matsumoto R., Mineshige S., 2001, PASJ, 53, L1

\bibitem{}
Matt G., Perola G.C., 1992, MNRAS, 259, 433

\bibitem{}
Miller J.M. et al., 2002, ApJ, 578, 348

\bibitem{}
Miller K.A., Stone J.M., 2000, ApJ, 534, 398

\bibitem{}
Nandra K., George I.M., Mushotzky R.F., Turner T.J., Yaqoob, T., 1997, 
ApJ, 477, 602

\bibitem{}
Novikov I.D., Thorne K.S., 1973, in Black Holes, eds. C. DeWitte \& B.S. DeWitte 
(New York: Gordon and Breach), p.~344

\bibitem{}
Onken C.A., Peterson B.M., Dietrich M., Robinson A., Salamanca I.M., 2003, 
ApJ, in press, astro-ph/0212115

\bibitem{}
Paczynski B., Wiita P.J., 1980, A\&A, 88, 23

\bibitem{}
Page D.N., Thorne K.S., 1974, ApJ, 191, 499

\bibitem{}
Papaloizou J.C.B., Nelson R.P., 2003, MNRAS, in press, astro-ph/0211493

\bibitem{}
Rauch K.P., Blandford R.D., 1994, ApJ, 421, 46

\bibitem{}
Remillard R.A., Morgan E.H., McClintock J.E., Bailyn C.D., Orosz J.A., 1999, 
ApJ, 522, 397

\bibitem{}
Reynolds C.S., Armitage P.J., 2001, ApJ, 561, L81

\bibitem{}
Reynolds C.S., Nowak M.A., 2003, Physics Reports, in press, astro-ph/0212065

\bibitem{}
Reynolds C.S., Young A.J., Begelman M.C., Fabian A.C., 1999, ApJ, 514, 164

\bibitem{}
Ruszkowski M., 2000, MNRAS, 315, 1

\bibitem{}
Steinacker A., Papaloizou J.C.B., 2002, ApJ, 571, 413

\bibitem{}
Stella L., 1990, Nature, 344, 747

\bibitem{}
Stone J.M., Norman M.L., 1992a, ApJS, 80, 753

\bibitem{}
Stone J.M., Norman M.L., 1992b, ApJS, 80, 791

\bibitem{}
Strohmayer T.E., 2001, ApJ, 554, L169

\bibitem{}
Tanaka Y. et al., 1995, Nature, 375, 659

\bibitem{}
Turner T.J. et al., 2002, ApJ, 574, L123

\bibitem{}
van der Klis M., 2000, ARA\&A, 38, 717

\bibitem{}
Wilms J., Reynolds C.S., Begelman M.C., Reeves J., 
Molendi S., Staubert R., Kendziorra E., 2001, MNRAS, 328, L27

\bibitem{}
Winters W.F., Balbus S.A., Hawley J.F., 2003, MNRAS, in press, astro-ph/0301498

\bibitem{}
Wu X.-B., Han J.L., 2001, ApJ, 561, L59

\end{thebibliography}
\end{document}